# Measurement of breast-tissue x-ray attenuation by spectral imaging: fresh and fixed normal and malignant tissue

Erik Fredenberg,[1,2] Paula Willsher,[3] Elin Moa,[2]
David R Dance,[4,5] Kenneth C Young[4,5] and Matthew G Wallis[3,6]

[1] Philips Research, Knarrarnäsgatan 7, 164 85 Kista, Sweden
[2] Philips Health Systems, Mammography Solutions, Torshamnsgatan 30A, 164 40 Kista, Sweden
[3] Cambridge Breast Unit, Addenbrookes Hospital, Hills Road, Cambridge CB2 0QQ, United Kingdom
[4] NCCPM, Royal Surrey County Hospital, Guildford GU2 7XX, United Kingdom
[5] Department of Physics, University of Surrey, Guildford GU2 7XH, United Kingdom
[6] NIHR Cambridge Biomedical Research Centre, Addenbrookes Hospital, Hills Road, Cambridge CB2 0QQ, United Kingdom

E-mail: erik.fredenberg@philips.com

**Abstract**
Knowledge of x-ray attenuation is essential for developing and evaluating x-ray imaging technologies. In mammography, measurement of breast density, dose estimation, and differentiation between cysts and solid tumours are example applications requiring accurate data on tissue attenuation. Published attenuation data are, however, sparse and cover a relatively wide range. To supplement available data we have previously measured the attenuation of cyst fluid and solid lesions using photon-counting spectral mammography. The present study aims to measure the attenuation of normal adipose and glandular tissue, and to measure the effect of formalin fixation, a major uncertainty in published data. A total of 27 tumour specimens, 7 fibro-glandular tissue specimens, and 15 adipose tissue specimens were included. Spectral (energy-resolved) images of the samples were acquired and the image signal was mapped to equivalent thicknesses of two known reference materials, from which x-ray attenuation as a function of energy can be derived. The spread in attenuation between samples was relatively large, partly because of natural variation. The variation of malignant and glandular tissue was similar, whereas that of adipose tissue was lower. Formalin fixation slightly altered the attenuation of malignant and glandular tissue, whereas the attenuation of adipose tissue was not significantly affected. The difference in attenuation between fresh tumour tissue and cyst fluid was smaller than has previously been measured for fixed tissue, but the difference was still significant and discrimination of these two tissue types is still possible. The difference between glandular and malignant tissue was close-to significant; it is reasonable to expect a significant difference with a larger set of samples. We believe that our studies have contributed to lower the overall uncertainty of breast tissue attenuation in the literature due to the relatively large sample sets, the novel measurement method, and by clarifying the difference between fresh and fixed tissue.

## 1. Introduction

Basic knowledge of the x-ray attenuation of tissue is essential for the development of new x-ray imaging technologies as well as for the study of existing technologies. Unenhanced spectral imaging, for example, is an emerging x-ray imaging technology that measures tissue properties, without injection of a contrast agent, using differences in attenuation as a function of energy (Fredenberg 2018).

In the field of mammography, efforts have been made to employ unenhanced spectral imaging to improve tumour visibility, but this exercise has so far proven difficult to realize because of the small difference in linear attenuation between malignant tissue and normal glandular tissue (Johns and Yaffe 1987). It may, however, be less complicated to differentiate between cyst fluid and solid tissue. Round lesions are common, and it can be difficult to distinguish between benign cysts and potentially malignant solid lesions using conventional x-ray imaging, particularly when the margin is partly obscured. Currently, recall rates from screening mammography are approximately 5% in Northern Europe and much higher in the United States (Smith-Bindman *et al* 2003), which is not only costly for the screening programme (Guerriero *et al* 2011), but also stressful for patients (Brett and Austoker 2001). Spectral imaging could potentially characterize a finding as cystic or solid in the screening image (Erhard *et al* 2016, Fredenberg *et al* 2013, 2016), thereby lowering the number of recalls and addressing the relatively low specificity of x-ray mammography screening.

Another application of unenhanced spectral imaging is measurement of the volumetric breast density (Ding and Molloi 2012, Johansson *et al* 2017), i.e., the fraction of fibro-glandular tissue in the breast. Accurate measures of breast density have the potential to improve risk assessments (Tice *et al* 2008), which in turn facilitates the transformation from population-based to stratified screening in order to increase sensitivity and minimize the cost of screening (Schousboe *et al* 2011). Breast density also provides important input to dose calculations (Dance and Sechopoulos 2016), and may be employed as a biomarker in treatment monitoring (Li *et al* 2013).

Accurate tissue attenuation data is crucial to the development of spectral x-ray techniques, such as lesion characterization and breast-density measurement. Accordingly, we have developed a method to measure the energy-dependent x-ray attenuation of tissue samples, which uses spectral imaging to map tissue attenuation to equivalent thicknesses of two reference materials. The method has previously been applied to measure the attenuation of cyst fluid and formalin-fixed solid breast lesion specimens (Fredenberg *et al* 2013, 2016). These studies, which showed a significant separation between cyst-fluid and solid-tissue attenuation, marked the first two steps in our efforts to evaluate the feasibility of using unenhanced spectral imaging for lesion characterization and laid the groundwork for a clinical pilot study (Erhard *et al* 2016).

Our measurement of cyst-fluid attenuation was the first of its kind, and the study on solid lesions more than doubled the aggregated number of samples in the literature. Nevertheless, these studies focused on lesion attenuation and no samples of normal tissue were included, despite the fact that adipose and glandular tissues make up the bulk of the breast and their attenuation properties are therefore of the utmost importance. Preliminary investigations have shown that data available in the literature on normal tissue attenuation exhibit a relatively large spread (Fredenberg *et al* 2015). The same situation exists for tumour attenuation, and the spread may be attributed to the limited number of samples in each study, which limits the accuracy of the estimated expectation values, but also to the different measurement setups employed (Fredenberg *et al* 2016). We therefore see a need to measure the attenuation of normal tissue within the same framework as we have used on lesions in order to make meaningful comparisons possible. We also see a need to contribute more samples to the available literature in order to reduce the overall error of the mean.

Further, our previous study on solid tissue was conducted on formalin-fixed specimens, which constitutes a considerable uncertainty. To our knowledge, not more than a single study has investigated the effect of fixation on breast-tissue attenuation; Chen *et al* (2010) did not find any significant difference between fresh and fixed adipose tissue, but the authors did report significant effects on glandular and tumour tissue. Fredenberg *et al* (2016) raised worries that fixation therefore might impact the separation of solid lesions from cysts, but the number of samples in Chen *et al* (2010) was relatively small (six) and the measurements were conducted in a relatively narrow energy range (17-23 keV). We therefore see a need to measure and compare the attenuation of a larger number of fresh and fixed specimens under conditions that are similar to the clinical environment.

Published data that are useful to determine the energy-dependent x-ray attenuation can be broadly categorized into two groups: 1) measurements of the elemental composition, which, in combination with elemental attenuation coefficients, can be used to calculate the linear attenuation according to the mixture rule, and 2) direct measurements of the linear attenuation coefficients. The perhaps most widely spread study on elemental composition was published by Hammerstein *et al* (1979) in order to make dose estimates and included samples of normal adipose and glandular tissue. Woodard and White (1986) added measurements from a few additional specimens of glandular tissue to the Hammerstein data, and these results were later cited in ICRU report 46 (ICRU 1992). Poletti *et al* (2002) measured the elemental composition of normal breast tissue samples as an intermediate step to determine angular scattering distributions. Relatively large differences compared to the Hammerstein data were noted for adipose tissue, whereas the measurements on glandular tissue agreed better. Pioneering direct attenuation measurements were presented by Johns and Yaffe (1987), who measured the linear attenuation coefficient of normal and cancerous tissue using a broad x-ray spectrum and a spectroscopy detector. Later, Carroll *et al* (1994) and Chen *et al* (2010) instead used monochromatic synchrotron radiation. Tomal *et al* (2010) used an x-ray tube and a silicon monochromator to measure the linear attenuation of breast tissue.

The purpose of the current study is to fill the above mentioned gaps in previous studies on breast-tissue attenuation and is therefore twofold: 1) to measure the x-ray attenuation of normal breast tissue, and 2) to measure the effect of formalin fixation on x-ray attenuation. Our immediate need for such data is to pursue the development of spectral imaging applications, but data on tissue attenuation may also have general value for the scientific community as the sources of such data are sparse.

## 2. Materials and Methods

### 2.1. Background of spectral imaging

The measurement procedure has been generally described in our previous publications (Fredenberg *et al* 2013, 2016), but is summarized here for clarity, and any differences are provided.

All measurements were conducted on a Philips MicroDose SI, which is a photon-counting spectral mammography system in a scanning multi-slit geometry. The tungsten-target x-ray tube is filtered with aluminium, and images were acquired at 32 kV acceleration voltage and an exposure level of 40 mAs. The scanning multi-slit geometry of the MicroDose system rejects virtually all scattered radiation (Åslund *et al* 2006). A low-energy threshold in the electronics rejects virtually all electronic noise. An additional high-energy threshold was set close to 20 keV to sort detected photons according to energy into two bins with approximately equal size. The energy resolution at this energy level is approximately 5 keV (Fredenberg *et al* 2010). The images were sub-sampled from 50 µm to 100 µm pixel size to facilitate the additional image processing carried out within the study.

X-ray attenuation at mammographic x-ray energies is approximately made up of only two independent interaction effects, namely photoelectric absorption and scattering processes (Alvarez and Macovski 1976, Lehmann *et al* 1981, Johns and Yaffe 1987, Fredenberg 2018). Accordingly, a linear combination of any two materials of different and low atomic number can approximately simulate the energy-dependent attenuation of a third material of a given thickness,

$$t_{\text{sample}}\mu_{\text{sample}}(E) = t_1\mu_1(E) + t_2\mu_2(E). \qquad (1)$$

We call these materials reference materials, and if this relationship is assumed to hold exactly, then the associated normalized reference thicknesses $[t_1, t_2]/t_{\text{sample}}$ are unique descriptors of the energy dependent sample attenuation ($\mu_{\text{sample}}$) given the known attenuations of the reference materials ($\mu_1$ and $\mu_2$). In other words, the detected signal in an x-ray detector would be identical for a tissue sample and for the equivalent combination of reference materials, regardless of incident energy spectrum or detector response.

Our method to measure the energy-dependent x-ray attenuation of tissue samples builds on Eq. (1) using aluminium (Al) and polymethyl methacrylate (PMMA) as the two reference materials. In common with previous studies (Lehmann *et al* 1981, Johns and Yaffe 1987), it is useful to also express the equivalent Al and PMMA thicknesses in terms of the corresponding Al-PMMA vector with magnitude and angle given by

$$r = \sqrt{t_{\text{Al}}^2 + t_{\text{PMMA}}^2} \qquad \text{and} \qquad \theta = \tan^{-1}\left(\frac{t_{\text{Al}}}{t_{\text{PMMA}}}\right). \qquad (2)$$

The magnitude $r$ is directly proportional to the thickness and the density (specific weight) of the sample, whereas the angle $\theta$ is related to the attenuation energy dependence and the (effective) atomic number of the material, and is independent of sample thickness.

*2.2. Spectral attenuation measurements*

In total, 37 tissue samples of solid malignant breast lesions and 22 samples containing fibro-glandular and / or adipose breast tissue were considered for the study. Ethical approval was obtained to use samples from women from whom generic consent had been obtained prior to surgery. Immediately post-surgery the tissue was sliced, the fresh tissue slices were promptly obtained from pathology, specimen images were acquired with the spectral mammography system, and the samples were returned without delay to pathology for formalin fixation. The same samples were again obtained from pathology post fixation to acquire a second set of spectral specimen images. The imaging and measurement procedure was carried out by an experienced radiographer (PW).

A provisional assessment of sample eligibility was conducted by visual inspection of the tissue. For the lesion samples it was required that at least one slice had the solid breast lesion clearly visible on both surfaces in order to avoid interference by other tissue types. For the normal tissue it was required that one of the slices contained homogeneous tissue through the entire slice for a reasonably large connected area. Due to the logistical constraints of obtaining tissue from, and returning it to, the pathology laboratory cut up room without disruption or delaying processing, final sample eligibility could only be verified after imaging had been performed.

Samples were placed in a hollow PMMA cylinder and were gently compressed and flattened by twisting the threaded lid. The sample thickness ($t_{\text{sample}}$) was calculated as the mean of a measurement with a protractor on the lid and a calliper measurement. Figure 1 shows photographs and x-ray images of sample holder and specimens. Regions-of-interest (ROIs) corresponding to homogeneous areas of adipose, fibro-glandular, or malignant tissue were selected in each image. For tumour samples, one ROI was manually selected to cover part of the lesion and was automatically divided into four equal-sized sub ROIs in order to estimate inhomogeneities within the sample. For adipose and glandular tissue, homogeneous areas were in general scattered over the sample and four sub ROIs of varying size were instead manually selected at various locations over the sample. ROI selection was done by a medical physicist (EF) and confirmed or adjusted by an experienced radiologist (MGW). Care was taken to avoid microcalcifications in the ROIs. Each sample was oriented similarly for imaging of fresh and fixed tissue, and ROIs were placed at similar locations for the fresh and fixed tissue.

An Al and PMMA step wedge was positioned adjacent to the sample to provide a reference grid of thickness / material combinations. ROIs were automatically selected on each step of the step wedge by projecting the sample ROI in the scanning direction (Figure 1). The range of the step wedge was adapted to the sample thickness by placing thin PMMA plates and Al sheets (step-wedge plates) on top of the step wedge. Preliminary investigations showed that adipose tissue exhibits negative Al thicknesses in the Al-PMMA space. To compensate for the negative thicknesses, adipose samples were imaged with an appropriate number of thin Al sheets (sample plates) above the sample. The thicknesses of PMMA and Al in the reference grid were calculated as the differences between the total thicknesses at the step-wedge ROIs (the sum of step-wedge plates, step wedge, and mounting plates) and the total thicknesses at the sample ROI (the sum of all sample holder components, sample plates, and mounting plates). The increased traversed thickness at oblique incidence was taken into account for all calculations.

X-ray attenuation was measured by mapping the high- and low energy counts obtained from the sample ROI against the reference grid obtained from the step-wedge ROIs. Linear Delaunay triangulation in the log domain was used for interpolation in the reference grid. In a limited number of cases, the range of the step wedge was misplaced and extrapolation by a second-degree surface in the log domain was employed outside the reference grid. Four images with identical ROI selection were acquired of each sample, but because Fredenberg *et al* (2016) concluded that the variation between subsequent images is minimal, the four readings were averaged without further analysis in the current study.

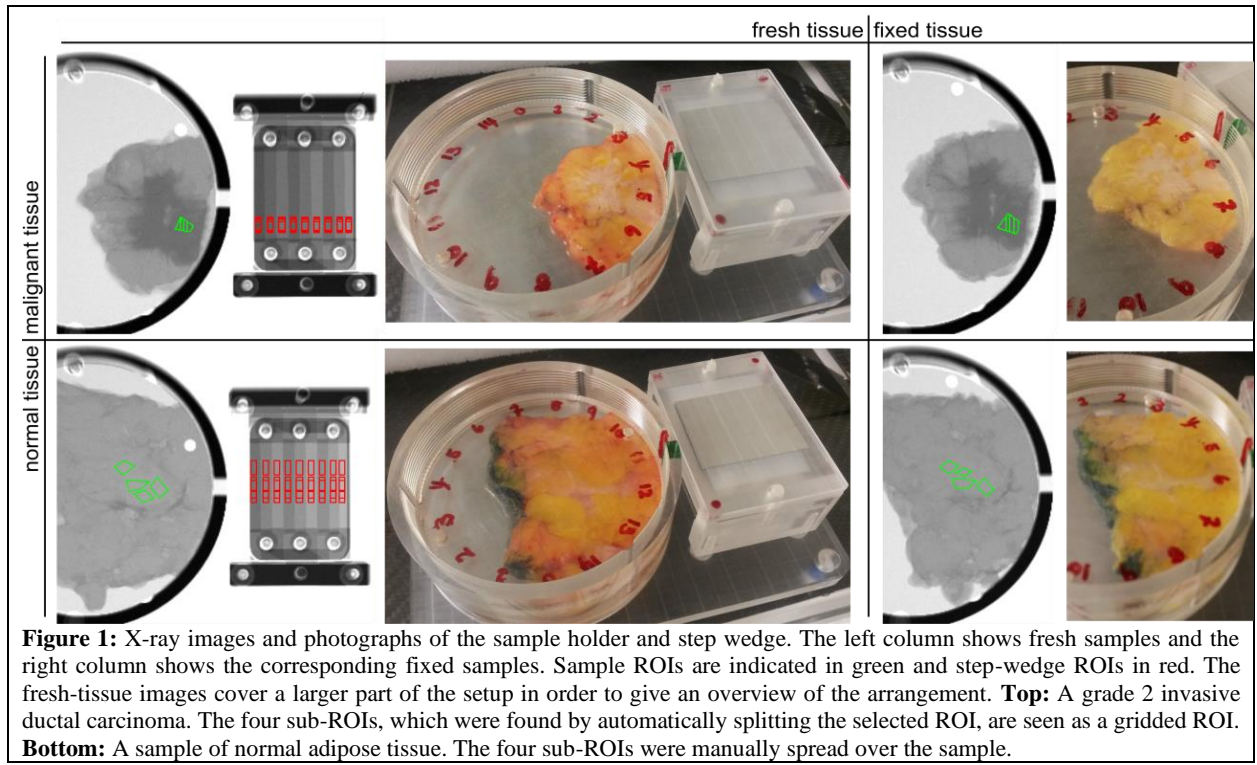

**Figure 1:** X-ray images and photographs of the sample holder and step wedge. The left column shows fresh samples and the right column shows the corresponding fixed samples. Sample ROIs are indicated in green and step-wedge ROIs in red. The fresh-tissue images cover a larger part of the setup in order to give an overview of the arrangement. **Top:** A grade 2 invasive ductal carcinoma. The four sub-ROIs, which were found by automatically splitting the selected ROI, are seen as a gridded ROI. **Bottom:** A sample of normal adipose tissue. The four sub-ROIs were manually spread over the sample.

## 2.3. Measurement errors and variabilities

For tumour tissue samples, the variation between the sub ROIs (referred to as the intra-image variability, $\sigma_{\text{intra I}}$) can be assumed to be caused by quantum noise ($\sigma_{\text{quant I}}$), random measurement errors from position-dependent thickness variations of the measurement setup and the sample ($\sigma_{\text{th I}}$), and intra-sample tissue variations ($\sigma_{\text{intra S}}$):

$$\sigma_{\text{intra I}}^2 = \sigma_{\text{intra S}}^2 + \sigma_{\text{th I}}^2 + \sigma_{\text{quant I}}^2, \tag{3}$$

where $\sigma_{\text{quant I}}$ was estimated by error propagation through the interpolation process, assuming knowledge of the photon counts per pixel and a Poisson distribution. Compared to Fredenberg *et al* (2016), $\sigma_{\text{quant I}}$ was a factor of two lower because the four readings of each sample were averaged in the present study.

The maximum error caused by linear interpolation of the slightly non-linear reference grid was estimated as the maximum deviation between the measured grid and a linear function extending from the periphery of the grid, divided over the edges of the grid. The expected variability caused by this error was estimated as the standard deviation of the rectangular distribution ranging from zero to the maximum error. Preliminary investigations showed that this variability on average amounted to less than 5 % of the quantum noise and the variability contribution from interpolation error was therefore ignored.

Fredenberg *et al* (2016) concluded that, for tumour tissue, inhomogeneity within an ROI was not a complete measure of sample inhomogeneity because the sub ROIs were located next to each other and covered only a portion of the lesion. In fact, no significant difference was found between $\sigma_{\text{intra I}}$ and $\sigma_{\text{quant I}}$ (i.e., $\sigma_{\text{intra S}}^2 + \sigma_{\text{th I}}^2 \approx 0$ for small ROIs). It is, however, likely that $\sigma_{\text{intra S}}$ and $\sigma_{\text{th I}}$ do not represent normally distributed stationary random processes; low spatial-frequencies typically dominate in the thickness variation of PMMA, Al, and sample, and the same is generally true for breast tissue contrast (Cederström and Fredenberg 2014), which means that $\sigma_{\text{th I}}$ and $\sigma_{\text{intra S}}$ can be expected to increase with distance between the sub ROIs. Hence, for the measurements on adipose / glandular tissue, these variabilities may contribute more to $\sigma_{\text{intra I}}$ because in that case the four sub ROIs were spread out over the sample, as opposed to tumour tissue for which sub ROIs were located next to each other. Therefore, to estimate $\sigma_{\text{th I}}$ for the measurements on adipose and glandular tissue, the thickness variations across the full area of the

PMMA and Al components were measured, and a sample thickness variation equal to the uncertainty of the sample thickness measurement was assumed. These variations were propagated through the interpolation process.

To estimate the expectation value for each sample of tumour tissue or normal adipose / glandular tissue, data from all sub ROIs were averaged. The spread between these values is referred to as the total variability:

$$\sigma_{tot}^2 = \sigma_{inter\ S}^2 + \sigma_{th\ S}^2 + \frac{1}{4} \times \sigma_{intra\ I}^2, \quad (4)$$

where $\sigma_{inter\ S}$ is natural tissue variation between samples and $\sigma_{th\ S}$ is the error caused by the per-sample thickness measurement. For tumour tissue, $\sigma_{th\ S}$ also includes thickness variations of the PMMA and Al components as these are not captured by $\sigma_{th\ I}$ without spacing between the sub ROIs. It is assumed that intra-sample tissue variations ($\sigma_{intra\ I}$) are reduced by averaging the individual ROI measurements. Accordingly, we define $\sigma_{quant\ S} \equiv 1/2 \times \sigma_{quant\ I}$ to simplify notation in the following.

The aggregated variability measures ($\sigma_{intra\ I}$ and $\sigma_{tot}$) were estimated as the standard deviation of measured data. As the number of samples were limited, we investigated the possibility to average the variabilities of fresh and fixed tissue in order to improve statistical power. We do not expect that the fixation process would affect variability, but the differences in total variability between fresh and fixed tissue were in any case tested for significance prior to averaging. As described above, quantum noise ($\sigma_{quant\ I}$ and $\sigma_{quant\ S}$) was estimated on a per-measurement basis from the detected number of counts, and variabilities caused by thickness variations ($\sigma_{th\ I}$ and $\sigma_{th\ S}$) were estimated based on expected thickness errors for an average sample or ROI. For more information about the procedures in estimating $\sigma_{quant}$ and $\sigma_{th}$ we refer to our previous study (Fredenberg *et al* 2016).

Natural variation of tissue ($\sigma_{intra\ S}$ and $\sigma_{inter\ S}$) was not estimated directly, but could to some extent be inferred from Eq. (3) and Eq. (4). With $\sigma_{intra\ S}$, $\sigma_{th\ I}$, and likely also $\sigma_{inter\ S}$ representing non-stationary, non-Gaussian processes, the validity of Eq. (3) and Eq. (4) is, however, limited; an additional assumption on equal ROI size and position is necessary. In fact, it is possible that the intra-sample ($\sigma_{intra\ S}$) and inter-sample ($\sigma_{inter\ S}$) variabilities are both caused by the same type of natural variation and that $\sigma_{intra\ S} \rightarrow \sigma_{inter\ S}$ as ROI size $\rightarrow \infty$. A more thorough analysis of these variabilities is, however, not within the scope of the current study. All variabilities and random measurement errors discussed above are summarized in Table 1.

**Table 1:** Summary of variabilities and random measurement errors discussed in Sec. 2.3.

| Intra-image (ROI-to-ROI) variabilities | | Inter-sample (sample-to-sample) variabilities | |
|---|---|---|---|
| $\sigma_{intra\ I}$ | intra-image variability, total variation between ROIs | $\sigma_{tot}$ | total variability, variation between samples |
| $\sigma_{quant\ I}$ | quantum noise between ROIs | $\sigma_{quant\ S}$ | quantum noise between samples |
| $\sigma_{th\ I}$ | thickness variations between ROIs, caused by unevenness in measurement setup and sample | $\sigma_{th\ S}$ | per-sample variations in the thickness measurement and in the measurement setup |
| $\sigma_{intra\ S}$ | intra-sample tissue variations | $\sigma_{inter\ S}$ | natural tissue variation between samples |

In addition to the random measurement errors related to measurement precision ($\sigma_{th\ I}$, $\sigma_{th\ S}$, $\sigma_{quant}$), we can expect systematic errors, related to measurement accuracy, caused by uncertainties in thickness and density that are constant for all measurements. We have estimated these systematic errors by propagation of thickness and density uncertainties according to Fredenberg et al (2016).

## 3. Results

Of the original 37 samples of solid malignant tissue, 10 samples were excluded from the study: 1 case was a DCIS with no mass lesion discernible on the images, 2 cases were benign fibroadenoma, and 7 cases were excluded for technical reasons, mainly because it was detected that tumour was not histologically present at both cut surfaces despite initial appearances of the fixed specimen block. Of the remaining 27 biopsy proven solid tissue specimens, 19 were invasive ductal carcinoma (3 grade I, 10 grade II, 6 grade III), 7 were lobular carcinoma (6 grade II, 1 grade III), and 1 was a grade I mucinous carcinoma. Of the original 22 samples of normal breast tissue, 4 samples were excluded from the study for technical reasons, mainly because no homogeneous area of either adipose or fibro-glandular tissue could be found within the sample. Of the remaining 18 samples, 15 were used for measurements on adipose tissue and 7 were used for measurements on fibro-glandular tissue. The relatively large number of exclusions is due to the fact that final sample eligibility could only be verified after imaging had been performed, as discussed in Sec. 2.2.

The mean sample thicknesses, including fresh and fixed samples, were 6.8 mm (malignant tissue), 6.0 mm (glandular tissue), and 6.2 mm (adipose tissue). The mean total ROI sizes (sum of four sub ROIs), including fresh and fixed samples, were 89 mm$^2$ (malignant tissue), 486 mm$^2$ (glandular tissue), and 227 mm$^2$ (adipose tissue). The substantially larger ROIs for adipose and glandular tissue were possible because of several large homogeneous areas in the samples over which the four sub ROIs could be spread, as opposed to the lesion samples where the malignant tissue was typically at a single confined location.

The systematic error caused by uncertainties in thickness and density of the measurement setup was estimated (c.f. Sec. 2.3) as 1-2 % and 1 % for an average sample in the Al- and PMMA-equivalent thicknesses, respectively. The variability caused by variations in thickness over the measurement area and random errors in the sample thickness measurements ($\sigma_{\text{th S}}$ and $\sigma_{\text{th I}}$) was estimated for an average sample to constitute 1-2 % and 0.5-1 % in the Al- and PMMA-equivalent thicknesses, respectively.

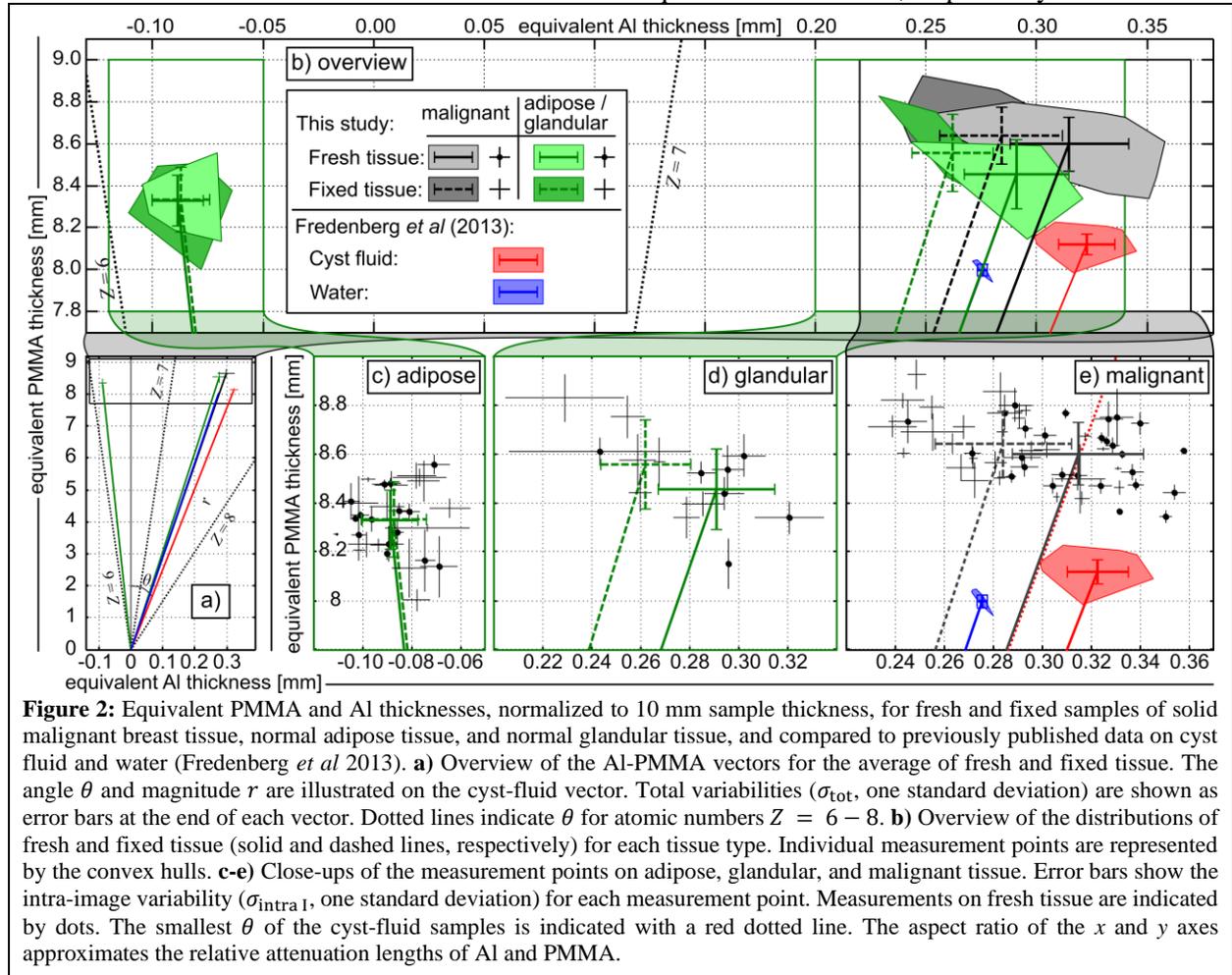

**Figure 2:** Equivalent PMMA and Al thicknesses, normalized to 10 mm sample thickness, for fresh and fixed samples of solid malignant breast tissue, normal adipose tissue, and normal glandular tissue, and compared to previously published data on cyst fluid and water (Fredenberg *et al* 2013). **a)** Overview of the Al-PMMA vectors for the average of fresh and fixed tissue. The angle $\theta$ and magnitude $r$ are illustrated on the cyst-fluid vector. Total variabilities ($\sigma_{\text{tot}}$, one standard deviation) are shown as error bars at the end of each vector. Dotted lines indicate $\theta$ for atomic numbers $Z = 6 - 8$. **b)** Overview of the distributions of fresh and fixed tissue (solid and dashed lines, respectively) for each tissue type. Individual measurement points are represented by the convex hulls. **c-e)** Close-ups of the measurement points on adipose, glandular, and malignant tissue. Error bars show the intra-image variability ($\sigma_{\text{intra I}}$, one standard deviation) for each measurement point. Measurements on fresh tissue are indicated by dots. The smallest $\theta$ of the cyst-fluid samples is indicated with a red dotted line. The aspect ratio of the *x* and *y* axes approximates the relative attenuation lengths of Al and PMMA.

Figure 2 shows measurement results from this study, as well as for 50 samples of cyst fluid and 50 samples of water from Fredenberg *et al* (2013), all expressed in terms of the equivalent Al and PMMA thicknesses, normalized to a sample thickness of 10 mm and a PMMA density of 1.19 g/cm$^3$. Figure 2a) shows an overview of the Al-PMMA vectors (average of fresh and fixed tissue) with the total variability ($\sigma_{\text{tot}}$) indicated as error bars at the end of each vector. The angle ($\theta$) and magnitude ($r$) are illustrated on the cyst-fluid vector. The (effective) atomic number of the sample determines $\theta$, and the angles for $Z = 6 - 8$ are indicated for illustration. Figure 2b) shows mean values and total variabilities ($\sigma_{tot}$) for each sample type as bold error bars, connected to the respective Al-PMMA vector, for fresh and fixed tissue. The individual measurement points are represented by their perimeters (convex hulls). Figure 2c-e) show close-ups of the measurements on adipose tissue, glandular, and malignant tissue. Individual measurement points are shown with error bars for the intra-image variability ($\sigma_{intra\ I}$). Measurements on fresh tissue are distinguished by dots. The red dotted line in Figure 2e) indicates the smallest $\theta$ of the cyst-fluid samples. In common with previous studies (Fredenberg *et al* 2016, Tomal *et al* 2010), we have bundled the data for malignant lesions of different type and grade. The aspect ratio of the *x* and *y* axes in Figure 2 (1:10) corresponds approximately to the relative attenuation lengths of Al and PMMA at mammography energies.

Table 2 lists all mean values and variability measures from the present study and from our previous study on solid tissue (Fredenberg *et al* 2016). Comparing present to previous data, there was a slight discrepancy in the mean values of fixed malignant tissue, but the difference was not significant (P > 0.05, two-sample *t*-test) and there is also no reason to expect a discrepancy. It therefore makes sense to attribute the difference to the large sample-to-sample variability, and the measures for a combined set of 83 samples in total are also listed in Table 2. In the following analysis we will, however, include only the current results, which have matching fresh tissue, but the larger data set on fixed tissue can be expected to have higher statistical power and may be useful outside this study. Further comparing the present data and our previous study, there was a relatively large and significant reduction in the total variability of PMMA thicknesses and $r$ (42 % reduction, P < 0.01, two-sample *F*-test). The differences in total variability of Al thicknesses and $\theta$ were, however, smaller and not significant (P > 0.3).

**Table 2:** Equivalent PMMA and Al thicknesses, normalized to 10 mm sample thickness, and the angle ($\theta$) and magnitude ($r$) of the Al-PMMA vectors. Data from the following sample sets are listed: Fresh and fixed solid malignant, glandular, and adipose tissue from the present study (sample set: this), fixed malignant tissue from Fredenberg *et al* (2016) (sample set: 2016), and all fixed malignant tissue combined into a single set (sample set: comb.). The following information is given for each sample set and measure: The number of samples (*n*); fixation status (fix.); mean value of the measure; total (sample-to-sample) variability of the measure ($\sigma_{tot}$, one standard deviation) with the expected quantum-noise contribution ($\sigma_{quant\ S}$); intra-image (sub ROI) variability ($\sigma_{intra\ I}$, one standard deviation).

| Sample type | set | *n* | fix. | PMMA thickness [mm] | | | | Al thickness [mm] | | | |
|---|---|---|---|---|---|---|---|---|---|---|---|
| | | | | mean | $\sigma_{tot}$ | $\sigma_{quant\ S}$ | $\sigma_{intra\ I}$ | mean | $\sigma_{tot}$ | $\sigma_{quant\ S}$ | $\sigma_{intra\ I}$ |
| malignant: | this | 27 | | 8.60 | 0.13 | 0.02 | 0.08 | 0.315 | 0.027 | 0.002 | 0.008 |
| | | | x | 8.64 | | | | 0.284 | | | |
| | 2016 | 56 | x | 8.59 | 0.22 | 0.05 | 0.09 | 0.296 | 0.028 | 0.004 | 0.009 |
| | comb. | 83 | x | 8.61 | 0.20 | 0.04 | 0.09 | 0.292 | 0.028 | 0.004 | 0.009 |
| glandular: | this | 7 | | 8.45 | 0.17 | 0.02 | 0.16 | 0.291 | 0.021 | 0.002 | 0.020 |
| | | | x | 8.56 | | | | 0.262 | | | |
| adipose: | this | 15 | | 8.33 | 0.14 | 0.02 | 0.10 | - 0.089 | 0.012 | 0.002 | 0.011 |
| | | | x | 8.33 | | | | - 0.087 | | | |

| Sample type | set | *n* | fix. | Magnitude ($r$) [mm] | | | | Angle ($\theta$) [mrad] | | | |
|---|---|---|---|---|---|---|---|---|---|---|---|
| | | | | mean | $\sigma_{tot}$ | $\sigma_{quant\ S}$ | $\sigma_{intra\ I}$ | mean | $\sigma_{tot}$ | $\sigma_{quant\ S}$ | $\sigma_{intra\ I}$ |
| malignant: | this | 27 | | 8.60 | 0.13 | 0.02 | 0.08 | 36.6 | 3.2 | 0.3 | 1.0 |
| | | | x | 8.64 | | | | 32.9 | | | |
| | 2016 | 56 | x | 8.59 | 0.22 | 0.05 | 0.09 | 34.5 | 3.4 | 0.6 | 1.1 |
| | comb. | 83 | x | 8.61 | 0.20 | 0.04 | 0.09 | 33.9 | 3.4 | 0.5 | 1.1 |
| glandular: | this | 7 | | 8.46 | 0.17 | 0.02 | 0.16 | 34.4 | 2.6 | 0.2 | 2.5 |
| | | | x | 8.56 | | | | 30.6 | | | |
| adipose: | this | 15 | | 8.33 | 0.14 | 0.02 | 0.10 | - 10.7 | 1.5 | 0.2 | 1.3 |
| | | | x | 8.33 | | | | - 10.5 | | | |

For malignant tissue, there were significant differences in attenuation between fresh and fixed tissue in terms of Al thickness and $\theta$ (P < 0.01, two-sample *t*-test), but the differences in PMMA thickness and $r$ were relatively smaller and not significant (P > 0.1). Compared to fixed lesions, the distribution of fresh lesions approached the previously published cyst-fluid distribution (the solid fresh-tissue vector in Figure 2e) is closer to the red cyst-fluid vector than is the dashed fixed-tissue vector), and the two tissue types were not significantly different in terms of Al thickness (P > 0.1, two-sample *t*-test), but differed significantly in terms of PMMA thickness, $\theta$, and $r$ (P < 0.01). For normal glandular tissue, there were, as for malignant tissue, significant average differences in attenuation between fresh and fixed tissue in terms of Al thickness and $\theta$ (P < 0.01, two-sample *t*-test), but not in terms of PMMA thickness or $r$ (P > 0.1). There was a consistent trend of all attenuation measures being slightly lower than for malignant tissue. If the measurements on fresh and fixed data for each sample are averaged (to increase power), this difference was close-to significant for all attenuation measures (P < 0.05, two-sample *t*-test). For normal adipose tissue, there was no significant difference between fresh and fixed tissue for any of the attenuation measures (P > 0.5, two-sample *t*-test), which is also manifested by the almost complete overlap of the mean values for fresh and fixed tissue in Figure 2c).

For all tissue types, the differences in total variability between fresh and fixed tissue were not significant (P > 0.3, two-sample *F*-test). As discussed in Sec. 2.3, the variability measures listed in Table 2 are therefore the averages of fresh and fixed tissue, and these average values were used in the following analysis of variability. For tumour tissue, the sample variability ($\sigma_{\text{intra I}}$) was significantly higher than the expected quantum noise between ROIs (P < 0.01, two-sample *t*-test between $\sigma_{\text{intra I}}$ and $\sigma_{\text{quant I}}$), in terms of PMMA and $r$ (53 % higher), and Al and $\theta$ (78 % higher on average). Note that $\sigma_{\text{quant I}}$ is not listed in Table 2 because the relationship to $\sigma_{\text{quant S}}$ stated in conjunction with Eq. (4) was found to hold, i.e. $\sigma_{\text{quant I}} = 2 \times \sigma_{\text{quant S}}$. For glandular tissue, the PMMA and $r$ sample variability was significantly higher than the expected quantum noise (341 % higher, P < 0.01, two-sample *t*-test), and a close-to significant difference was found between the Al and $\theta$ sample variability and the expected quantum noise (519 % higher on average, P < 0.05). For adipose tissue, the sample variability was significantly higher than the expected quantum noise (P < 0.01, two-sample *t*-test), in terms of PMMA and $r$ (152 % higher), and Al and $\theta$ (220 % higher on average).

For all tissue types, the total variability ($\sigma_{\text{tot}}$) was several times larger than the expected quantum noise (P < 0.01, chi-square variance test between $\sigma_{\text{tot}}$ and $\sigma_{\text{quant S}}$). There was no significant difference in total variability between malignant and glandular tissue (P > 0.1, two-sample *F*-test on the mean values). Malignant and glandular tissue exhibited significantly higher total variability than adipose tissue in terms of Al and $\theta$ (on average 118 % higher for tumour tissue and 71 % higher for glandular tissue, P < 0.01). No significant difference was found in PMMA and $r$ total variability between adipose tissue and the other tissue types (P > 0.10).

For reference and comparison to other studies, the linear attenuation coefficients of fresh and fixed malignant, adipose, and glandular breast tissue, calculated from the measured PMMA- and Al-equivalent thicknesses at a range of x-ray energies relevant to mammography, are shown in Table 3.

**Table 3:** Linear attenuation coefficients of malignant and normal glandular and adipose breast tissue calculated from the measured PMMA- and Al-equivalent thicknesses. The number of samples (*n*) and fixation status (fix.) are listed for each sample set. The total variability ($\sigma_{\text{tot}}$, one standard deviation) is given in parenthesis.

| Sample type | *n* | fix. | linear attenuation [cm$^{-1}$] | | | | | |
|---|---|---|---|---|---|---|---|---|
| | | | 15 keV | 20 keV | 25 keV | 30 keV | 35 keV | 40 keV |
| malignant: | 27 | | 1.80 (0.054) | 0.877 (0.023) | 0.551 (0.012) | 0.406 (0.008) | 0.332 (0.006) | 0.289 (0.004) |
| | | x | 1.74 (0.053) | 0.851 (0.023) | 0.538 (0.012) | 0.398 (0.007) | 0.326 (0.005) | 0.285 (0.004) |
| glandular: | 7 | | 1.73 (0.045) | 0.845 (0.020) | 0.532 (0.010) | 0.394 (0.007) | 0.322 (0.005) | 0.281 (0.004) |
| | | x | 1.69 (0.022) | 0.825 (0.009) | 0.523 (0.004) | 0.388 (0.003) | 0.319 (0.003) | 0.280 (0.003) |
| adipose: | 15 | | 0.902 (0.027) | 0.484 (0.012) | 0.338 (0.007) | 0.273 (0.005) | 0.239 (0.004) | 0.219 (0.004) |
| | | x | 0.906 (0.037) | 0.486 (0.017) | 0.339 (0.010) | 0.274 (0.007) | 0.240 (0.006) | 0.220 (0.005) |

## 4. Discussion

*4.1. Measurement variability*

As we have noted previously (Fredenberg *et al* 2016), the total variability ($\sigma_{\text{tot}}$) of tumour samples is substantially larger than what would be expected from measurement errors (i.e., thickness errors of the measurement setup – $\sigma_{\text{th S}}$, $\sigma_{\text{th I}}$ – and quantum noise – $\sigma_{\text{quant S}}$). Referring to Eq. (4), a large part of $\sigma_{\text{tot}}$ for tumour tissue can therefore be explained by natural variation ($\sigma_{\text{inter S}}$ and $\sigma_{\text{intra S}}$). In particular the Al variability is more strongly associated with natural variation (random measurement errors constitute approximately 10 % of $\sigma_{\text{tot}}$), whereas the PMMA variability contains a larger component of measurement errors (approximately 30 %). This difference can be understood by modelling the natural variation of tumour tissue as a blending with other tissue types (in particular glandular tissue); considering the span of tissue attenuation in the present study we see that the spread is mainly in the Al direction. As a consequence of the steep $\theta$, the behaviour of $r$ variability follows approximately that of PMMA and the $\theta$ variability follows Al.

For glandular tissue, the total variability was not significantly different from that of tumour tissue and the contribution by random measurement errors was approximately the same in Al thicknesses (10%). For adipose tissue on the other hand, the total variability in Al thicknesses was significantly lower than for any of the other tissue types. Random measurement errors constitute approximately 15% of the variability in Al, which is higher than for glandular and tumour tissue. As discussed above, Al thicknesses have a stronger association to natural variation and these results therefore indicate that adipose tissue has the lowest natural variation of the investigated tissues. The high homogeneity of the adipose tissue samples was also corroborated by visual inspection; the adipose samples appeared more homogeneous on the x-ray images.

Compared to our previous study on solid malignant breast tissue, $\sigma_{\text{tot}}$ was significantly and substantially lower in terms of PMMA thickness and $r$, but not in terms of Al thickness or $\theta$. Following the discussion above, the reduction in total variability in PMMA and $r$ implies a reduction in measurement errors, which makes sense because there is no reason to expect a reduction in natural tissue variation. The average sample thicknesses were approximately the same in both studies (6.8 mm vs. 6.4 mm), but because of larger ROI sizes in the present study (89 mm$^2$ vs. 54 mm$^2$), the quantum noise was lower, which may account for part of the reduction in $\sigma_{\text{tot}}$. Further, even though the measurement setups were identical in the two studies, the manual thickness measurement of the sample might be associated with a learning curve so that the precision has improved over time (lower $\sigma_{\text{th S}}$). Another learning curve might be coupled to the tissue handling, leading to less unevenness of the sample (lower $\sigma_{\text{th I}}$).

Another difference compared to our previous results is that, in the present study, $\sigma_{\text{intra I}}$ for tumour tissue was significantly larger than $\sigma_{\text{quant I}}$, whereas previously no significant difference was found. The reason for this discrepancy may be the larger ROI sizes, which together with averaging of the four readings reduces the quantum noise so that intra-image variations ($\sigma_{\text{intra S}}$ and $\sigma_{\text{th I}}$) become more visible. Further, as noted in Sec. 2.3, it is likely that $\sigma_{\text{intra S}}$ and $\sigma_{\text{th I}}$ grow with distance between the sub ROIs and will therefore be higher for the larger ROI sizes in the present study. For glandular and adipose tissue, the ROI sizes were larger than for tumour tissue, which certainly resulted in lower quantum noise, but as an effect of the spreading of the four sub ROIs, random fluctuations within each sample ($\sigma_{\text{intra I}}$) were substantially higher than for tumour tissue and constituted a substantially larger portion of $\sigma_{\text{tot}}$.

*4.2. Discrimination between tissue types*

Roughly speaking, x-ray attenuation is determined by the effective atomic number, density (specific weight), and thickness of the attenuating material. Spectral imaging enables differentiation between atomic number and density by measuring the energy dependence of the attenuation (Fredenberg 2018). It may therefore be possible to discriminate between cyst fluid, glandular tissue, and solid malignant tissue despite the very small differences in linear attenuation (approximately 3% average difference in the 15-40 keV interval according to Table 3 and Fredenberg *et al* (2013), Table 4) because the energy dependencies are not the same. In terms of the Al-PMMA vector, spectral imaging enables measurement of the angle ($\theta$), which is related to the effective atomic number and independent of the sample thickness, whereas the magnitude ($r$) is directly proportional to both the density and the thickness of the sample and therefore cannot be used for material discrimination unless the thickness is known. In clinical applications, the thickness may be available from 3D information or from assumptions of slow thickness variations, but it may also be an unknown variable, in which case $\theta$ can be used for discrimination.

Fredenberg *et al* (2016) concluded that the attenuation of cyst fluid and fixed tumour specimens differed significantly in terms of $\theta$ and $r$, but one concern raised was that formalin-fixed tissue was used as opposed to fresh tissue, which is a potential bias when making direct interpretation of the results to a clinical situation. The present measurements on fresh tumour tissue showed a smaller, although still significant, difference compared to cyst fluid in terms of $\theta$ and $r$. Hence, discrimination of these two tissue types in clinical practice may be more challenging than previously believed, but is still possible as also evidenced by a clinical pilot study conducted by Erhard et al (2016). Despite the large spread, no solid samples fall within the shaded region of the cyst distribution (Figure 2), but 15 malignant samples (56% of the total number) overlap with the cyst distribution in terms of $\theta$ (Figure 2, area below the dotted line), and these samples would therefore be challenging to distinguish from cyst fluid.

We found a close-to significant difference between normal glandular tissue and malignant tissue in terms of all attenuation measures. Both $\theta$ and $r$ were slightly higher for malignant tissue, which implies higher effective atomic number and higher density than glandular tissue. This finding involved averaging over fresh and fixed tissue to increase statistical power. The result should therefore be treated with caution, but it is reasonable to expect a significant difference between the two tissue types with a larger set of glandular tissue samples. The difficulty to discriminate between the attenuation of glandular and malignant tissue is further underlined by mixed results in the literature; two previous studies have found significant differences between malignant and glandular tissue at lower energies (Johns and Yaffe 1987, Tomal *et al* 2010), whereas at least one other study was not able to show a significant difference (Chen *et al* 2010).

*4.3. Comparison to published data*

Figure 3 shows a comparison in terms of the equivalent Al and PMMA thicknesses between the data presented in this study and most of the attenuation data on breast tissue available in the literature. The mean values and total variabilities of the present study are represented by error bars, overlaid on the convex hulls of all measured data. Published studies are represented by filled (fresh) and open (fixed) markers, and the average of all studies, as well as the variation between mean values (one standard deviation), are shown as error bars. Mean values and variabilities are also listed in Table 4. In some cases, uncertainty measures are available from individual published studies, but these are not shown as variabilities in Figure 3 or Table 4 because no consistent way of converting linear attenuation uncertainty to Al and PMMA space could be found.

The studies included in Figure 3, Table 4, and the following analysis are: Hammerstein *et al* (1979), elemental composition of fresh tissue, 8 samples of adipose tissue, 5 samples of glandular tissue; Johns and Yaffe (1987), linear attenuation of fresh tissue in the energy range 18-110 keV, 7 samples of adipose tissue, 8 samples of fibrous tissue, 6 samples of infiltrating ductal carcinoma; Chen *et al* (2010), linear attenuation of fixed tissue in the energy range 15-26.5 keV (but all samples were not imaged at all energies), ≤13 samples of fibrous tissue, ≤17 samples of adipose tissue, ≤14 samples of cancerous tissue, mainly ductal carcinoma; Tomal *et al* (2010), linear attenuation coefficient of fixed tissue in the energy range 8-30 keV, 4 samples of glandular tissue, 28 samples of adipose tissue, 18 invasive ductal carcinomas, 6 fibroadenomas. Data from Woodard and White (1986), Carroll *et al* (1994), and Poletti *et al* (2002) were not included because the samples could not be unambiguously categorized as fresh or fixed.

For calculating the average of fixed tumour attenuation, we have treated our previous study (Fredenberg *et al* 2016) and the present one as separate studies, although the systematic errors are likely similar. The resulting higher weighting of our results can be motivated by the considerably larger set of samples (83 samples in the two studies) compared to other studies (~32 samples in total), which likely yields lower random errors. Equivalent Al and PMMA thicknesses for published data were found by fitting to linear attenuation coefficients according to Eq. (3) in Fredenberg *et al* (2016), under the conditions of 10 mm tissue thickness and 1.19 g/cm$^3$ PMMA density. The validity of this fit is explained in Fredenberg *et al* (2016). For attenuation data based on elemental composition, the linear attenuation was first calculated by the mixture rule and elemental attenuation from Berge*r et al* (2009).

The relatively large spread in the available data on tissue attenuation may be caused by 1) large natural spread between samples, 2) random measurement errors, and 3) different experimental conditions in the different studies that cause systematic differences and errors (Fredenberg *et al* 2016). Assuming that random fluctuations (uncertainties 1 and 2) are approximately the same in all studies and that the systematic errors (uncertainty 3) are random between studies, adding more studies to the literature will help reduce the error of the mean, and it is fair to assume that the mean values (error bars in Figure 3 and values in Table 4) represent better estimates of the expectation values than do the individual studies.

Chen *et al* (2010) measured the linear attenuation coefficients before and after formalin fixation for six of their samples, and the fitted Al- and PMMA-equivalent thicknesses are shown in Figure 3 and Table 4. The mean values for the subset of fixed tissue (cyan filled circles) do not in all cases agree with those of the full study (black or green filled circles), a discrepancy that may be caused by the use of a smaller number of samples and the narrower energy range (17-23 keV) compared to the full set of data. Data for the six fresh samples were therefore not included when calculating the mean of all studies. The differences between fixed tissue (Figure 3, cyan open circle) and fresh tissue (cyan filled circle) within the subset are, however, likely less prone to systematic errors as these can be expected to cancel when taking the difference. On the other hand, the differences are subject to random errors from both measurements, with the exception of sample-to-sample tissue variation ($\sigma_{\text{inter S}}$) as the same samples were used for measurements on both fresh and fixed tissue.

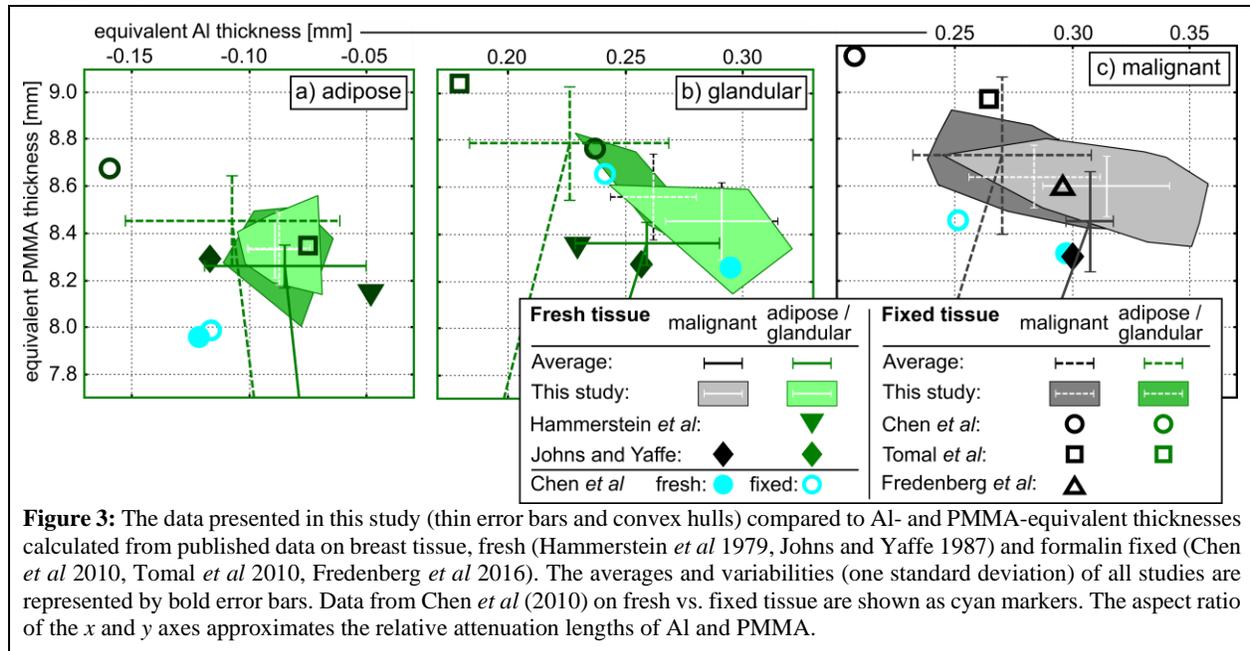

**Figure 3:** The data presented in this study (thin error bars and convex hulls) compared to Al- and PMMA-equivalent thicknesses calculated from published data on breast tissue, fresh (Hammerstein *et al* 1979, Johns and Yaffe 1987) and formalin fixed (Chen *et al* 2010, Tomal *et al* 2010, Fredenberg *et al* 2016). The averages and variabilities (one standard deviation) of all studies are represented by bold error bars. Data from Chen *et al* (2010) on fresh vs. fixed tissue are shown as cyan markers. The aspect ratio of the *x* and *y* axes approximates the relative attenuation lengths of Al and PMMA.

Similarly, the differences between fresh and fixed tissue measured in this study can be expected to contain a low level of systematic errors and, in addition, a lower level of random errors than the measurement by Chen *et al* (2010) because of the larger number of samples. The differences are listed in Table 4 along with the variability (the standard deviation over individual samples). The calculated differences between separate studies on fresh and fixed tissue are also listed in Table 4. Even though this calculation includes the largest number of samples, systematic errors and sample to sample variation can be expected to influence the results. In fact, the variability of the difference (propagated from the variability between studies) is in all cases larger than the variability for the present study.

For malignant tissue and glandular tissue, our results agree qualitatively with Chen *et al* (2010) as well as with the average of all studies, i.e., the fixation process yields higher equivalent PMMA thickness and lower equivalent Al thickness. For adipose tissue, Chen *et al* (2010) did not, in agreement with our study, find a significant difference between fresh and fixed tissue, which indicates low uptake of formalin in adipose tissue and an otherwise negligible effect of fixation. The difference between different publications on fresh and fixed adipose tissue is relatively large, but given the substantial variability in the estimation, the deviation is not significant.

## 5. Conclusions

The present study marks the next step in our efforts to characterize breast tissue attenuation. The series of studies now include cyst fluid and tumour tissue, as well as normal adipose and glandular tissue, all measured with close-to identical methods and in an environment relevant for screening mammography.

Solid samples exhibit large sample-to-sample variation compared to liquid samples and compared to what would be expected from measurement errors. We expect a major part of the variability, in particular for Al-equivalent thicknesses and $\theta$, to be caused by natural spread between samples. The natural variation of tumour tissue and glandular tissue is similar, whereas that of adipose tissue appears to be lower.

Fixation does not affect the attenuation of adipose tissue significantly, but yields higher equivalent PMMA thickness and lower equivalent Al thickness for malignant tissue and glandular tissue. Consequently, fresh tumour tissue showed a smaller, although still significant, difference compared to cyst fluid than has previously been measured for fixed tissue. Hence, discrimination of these two tissue types with spectral mammography may be more challenging than previously believed, but is still possible, as corroborated by a clinical pilot study conducted by Erhard *et al* (2016). The difference between normal glandular tissue and malignant tissue would likely be significant with a larger set of glandular tissue samples.

The present study contributes to lower the overall uncertainty in the literature by 1) a relatively large sample set, which reduces random errors and the effect of natural spread, 2) adding another measurement method, which helps reduce systematic uncertainty, and 3) clarifying the difference between fresh and formalin-fixed tissue, one of the major contributors to systematic discrepancies.

**Table 4:** Comparison of the equivalent PMMA and Al thicknesses, stratified on fresh and fixed tissue (fix.) and normalized to 10 mm tissue thickness, for the average of all published studies on each tissue type (including the present study), as measured by Chen *et al* (2010), and as measured in the present study. Mean values as well as the difference between mean values are given. For each set, the number of samples (Chen *et al*, this study), or the number of studies (all studies), are listed as *n*. Variability measures are given in parenthesis as the total variability (this study) or the variation between mean values (all studies). No variabilities are given for Chen *et al* (2010) because the published linear attenuation uncertainty could not be unambiguously converted to Al and PMMA.

| Sample type | Set | *n* | fix. | PMMA thickness [mm] mean | PMMA difference | Al thickness [mm] mean | Al difference |
|---|---|---|---|---|---|---|---|
| malignant: | all studies | 2 |   | 8.45 (0.21) | - 0.28 (0.40) | 0.308 (0.010) | 0.037 (0.039) |
|  |  | 3 | x | 8.73 (0.34) |  | 0.270 (0.038) |  |
|  | Chen *et al* | 6 |   | 8.31 | - 0.14 | 0.297 | 0.046 |
|  |  |  | x | 8.45 |  | 0.252 |  |
|  | this study | 27 |   | 8.60 (0.13) | - 0.04 (0.13) | 0.315 (0.027) | 0.031 (0.026) |
|  |  |  | x | 8.64 |  | 0.284 |  |
| glandular: | all studies | 3 |   | 8.36 (0.09) | - 0.42 (0.26) | 0.259 (0.031) | 0.033 (0.052) |
|  |  | 3 | x | 8.79 (0.24) |  | 0.226 (0.042) |  |
|  | Chen *et al* | 6 |   | 8.26 | - 0.20 | 0.295 | 0.043 |
|  |  |  | x | 8.65 |  | 0.241 |  |
|  | this study | 7 |   | 8.45 (0.17) | - 0.10 (0.15) | 0.291 (0.021) | 0.029 (0.010) |
|  |  |  | x | 8.56 |  | 0.262 |  |
| adipose: | all studies | 3 |   | 8.25 (0.08) | - 0.21 (0.21) | - 0.084 (0.034) | 0.023 (0.057) |
|  |  | 3 | x | 8.45 (0.19) |  | - 0.107 (0.046) |  |
|  | Chen *et al* | 6 |   | 7.96 | - 0.03 | - 0.121 | - 0.005 |
|  |  |  | x | 7.99 |  | - 0.116 |  |
|  | this study | 15 |   | 8.33 (0.14) | 0.00 (0.15) | - 0.089 (0.012) | 0.001 (0.011) |
|  |  |  | x | 8.33 |  | - 0.087 |  |

## Acknowledgements

Special thanks are extended to Dr E Provenzano, Breast Histopathologist, and Mr James Neal, Advanced Practitioner in Breast Dissection, both at the Department of Histopathology, Addenbrooke's Hospital and Cambridge NIHR Biomedical Research Centre, for their support in preparing the tissue samples. Part of this work was carried out within the OPTIMAM2 project funded by Cancer Research UK (grant number C30682/A17321). The Cambridge Human Research Tissue Bank is supported by the NIHR Cambridge Biomedical Research Centre.

# References

Alvarez R E and Macovski A 1976 Energy-selective reconstructions in X-ray computerized tomography. *Phys. Med. Biol.* **21** 733–44

Åslund M, Cederström B, Lundqvist M and Danielsson M 2006 Scatter rejection in multislit digital mammography *Med. Phys.* **33** 933–40

Berger M J, Hubbell J H, Seltzer S M, Coursey J S and Zucker D S 2009 XCOM: Photon Cross Section Database **2009** version 1.4 Online: http://physics.nist.gov/xcom

Brett J and Austoker J 2001 Women who are recalled for further investigation for breast screening: psychological consequences 3 years after recall and factors affecting re-attendance *J. Public Health Med.* **23** 292–300

Carroll F E, Waters J W, Andrews W W, Price R R, Pickens D R, Willcott R, Tompkins P, Roos C, Page D, Reed G, Ueda A, Bain R, Wang P and Bassinger M 1994 Attenuation of Monochromatic X-Rays by Normal and Abnormal Breast Tissues *Invest. Radiol.* **29** 266–72

Cederström B and Fredenberg E 2014 The influence of anatomical noise on optimal beam quality in mammography *Med. Phys.* **41** 121903

Chen R C, Longo R, Rigon L, Zanconati F, De Pellegrin A, Arfelli F, Dreossi D, Menk R-H H, Vallazza E, Xiao T Q and Castelli E 2010 Measurement of the linear attenuation coefficients of breast tissues by synchrotron radiation computed tomography. *Phys. Med. Biol.* **55** 4993–5005

Dance D R and Sechopoulos I 2016 Dosimetry in x-ray-based breast imaging *Phys. Med. Biol.* **61** R271

Ding H and Molloi S 2012 Quantification of breast density with spectral mammography based on a scanned multi-slit photon-counting detector: a feasibility study *Phys Med Biol* **57** 4719–38

Erhard K, Kilburn-Toppin F, Willsher P, Moa E, Fredenberg E, Wieberneit N, Buelow T and Wallis M G 2016 Characterization of Cystic Lesions by Spectral Mammography: Results of a Clinical Pilot Study *Invest. Radiol.* **51** 340–7

Fredenberg E 2018 Spectral and dual-energy X-ray imaging for medical applications *Nucl. Inst. Methods Phys. Res. A* **878** 74–87

Fredenberg E, Dance D R, Willsher P, Moa E, von Tiedemann M, Young K C and Wallis M G 2013 Measurement of breast-tissue x-ray attenuation by spectral mammography: first results on cyst fluid. *Phys. Med. Biol.* **58** 8609–20

Fredenberg E, Dance D R, Young K C, Moa E, Willsher P, Kilburn-Toppin F and Wallis M G 2015 X-ray attenuation of normal and cancerous breast tissue measured with photon- counting spectral imaging *Radiological Society of North America 2015 Scientific Assembly and Annual Meeting* Paper SSA 20-07 (Chicago, IL: RSNA)

Fredenberg E, Kilburn-Toppin F, Willsher P, Moa E, Danielsson M, Dance D R, Young K C and Wallis M G 2016 Measurement of breast-tissue x-ray attenuation by spectral mammography: solid lesions *Phys. Med. Biol.* **61** 2595–612

Fredenberg E, Lundqvist M, Cederström B, Åslund M and Danielsson M 2010 Energy resolution of a photon-counting silicon strip detector *Nucl. Instruments Methods Phys. Res. A* **613** 156–62

Guerriero C, Gillan M G C, Cairns J, Wallis M G and Gilbert F J 2011 Is computer aided detection (CAD) cost effective in screening mammography? A model based on the CADET II study *BMC Health Serv. Res.* **11**

Hammerstein G R, Miller D W, White D R, Masterson M E, Woodard H Q and Laughlin J S 1979 Absorbed Radiation Dose in Mammography *Radiology* **130** 485–91

ICRU 1992 *ICRU Report 46: Photon, electron, proton and neutron interaction data for body tissues* vol 46

Johansson H, von Tiedemann M, Erhard K, Heese H, Ding H, Molloi S and Fredenberg E 2017 Breast-density measurement using photon-counting spectral mammography *Med. Phys.* **44** 3579–93

Johns P C and Yaffe M J 1987 X-ray characterisation of normal and neoplastic breast tissues *Phys. Med. Biol.* **32** 675–95

Lehmann L A, Alvarez R E, Macovski A, Brody W R, Pelc N J, Riederer S J and Hall A L 1981 Generalized image combinations in dual KVP digital radiography *Med. Phys.* **8** 659–67

Li J, Humphreys K, Eriksson L, Edgren G, Czene K and Hall P 2013 Mammographic density reduction is a prognostic marker of response to adjuvant tamoxifen therapy in postmenopausal patients with breast cancer *J Clin Oncol* **31** 2249–56

Poletti M E, Goncalves O D and Mazzaro I 2002 X-ray scattering from human breast tissues and breast-equivalent materials *Phys. Med. Biol.* **47** 47–63

Schousboe J T, Kerlikowske K, Loh A and Cummings S R 2011 Personalizing mammography by breast density and other risk factors for breast cancer: analysis of health benefits and cost-effectiveness. *Ann. Intern. Med.* **155** 10–20

Smith-Bindman R, Chu P W, Miglioretti D L, Sickles E A, Blanks R, Ballard-Barbash R, Bobo J K, Lee N C, Wallis M G, Patnick J and Kerlikowske K 2003 Comparison of Screening Mammography in the United States and the United Kingdom *JAMA* **290** 2129–37

Tice J A, Cummings S R, Smith-Bindman R, Ichikawa L, Barlow W E and Kerlikowske K 2008 Using clinical factors and mammographic breast density to estimate breast cancer risk: development and validation of a new predictive model *Ann Intern Med* **148** 337–47

Tomal A, Mazarro I, Kakuno E M and Poletti M E 2010 Experimental determination of linear attenuation coefficient of normal, benign and malignant breast tissues *Radiat. Meas.* **45** 1055–9

Woodard H Q and White D R 1986 The composition of body tissues *Br. J. Radiol.* **59** 1209–19